# Scalar perturbations in $f(T)$ gravity using the 1 + 3 covariant approach

Shambel Sahlu[1,2,a], Joseph Ntahompagaze[1,3], Amare Abebe[4], Álvaro de la Cruz-Dombriz[5], David F. Mota[6]

[1] Astronomy and Astrophysics Research and Development Department, Ethiopian Space Science
  and Technology Institute, Addis Ababa, Ethiopia
[2] Department of Physics, College of Natural and Computational Science, Wolkite University, Wolkite , Ethiopia
[3] Department of Physics, College of Science and Technology, University of Rwanda, Kigali, Rwanda
[4] Center for Space Research, North-West University, Vanderbijlpark, South Africa
[5] Cosmology and Gravity Group, Mathematics and Applied Mathematics, University of Cape Town, Rondebosch 7701, South Africa
[6] Institute of Theoretical Astrophysics, University of Oslo, Oslo, Norway

**Abstract** The cosmological scalar perturbations of standard matter are investigated in the context of extended teleparallel $f(T)$ gravity theories using the 1 + 3 covariant formalism. After a review of the background gravitational field equations of $f(T)$ gravity and the introduction of the covariant perturbation variables, the usual scalar and harmonic decomposition have been performed, and the analysis of the growth of the density contrasts in the quasi-static approximation for two non-interacting fluids scenarios, namely torsion-dust and torsion-radiation mixtures is presented for the generic $f(T)$ gravity theory. Special applications to two classes of $f(T)$ gravity toy models, namely $f(T) = \mu T_0 \left( \frac{T}{T_0} \right)^n$ and $f(T) = T + \mu T_0 \left( -\frac{T}{T_0} \right)^n$, have then been made within the observationally viable regions of their respective parameter spaces, and the growth of the matter density contrast for both torsion-dust and torsion-radiation epochs of the Universe has been examined. The exact solutions of the dust perturbations, with growing amplitudes in cosmic time, are obtained for some limiting cases of $n$. Similarly, the long- and short-wavelength modes in the torsion-radiation case are treated, with the amplitudes either oscillating or monotonically growing with time. Overall, it is noted that $f(T)$ models contain a richer set of observationally viable structure growth scenarios that can be tested against up-and-coming observational data and can accommodate currently known features of the large-scale structure power spectrum in the general relativistic and $\Lambda CDM$ limits.

## 1 Introduction

The recent discovery of the accelerating expansion of the Universe [1,2] together with the anisotropy of the Cosmic Microwave Background Radiation (CMB) [3–5], and how cosmological perturbations [6–10] and the primordial fluctuations of the early Universe formed the large-scale structures [3,4,7,11] require to go beyond the standard model of cosmology. One of the modified gravity theories currently under active exploration is $f(T)$ gravity [12–20], where $T$ represents the torsion scalar. The $f(T)$ gravitational theory can resolve a number of longstanding issues in general relativity (GR), e.g., to study the primordial nucleosynthesis [21], the static spherically symmetric self-gravitating objects [22,23], observational constraints [24,25], the background and perturbation analysis in the metric formalism [26–31], just to mention a few. The study of linear cosmological perturbations in $f(T)$ gravity theory using the 1 + 3 covariant formalism is the main focus of research in this manuscript. Basically, there are two mainstream formalism to study cosmological perturbations, namely the metric formalism [8,32,33] and the 1 + 3 covariant gauge-invariant formalism [6,34–37], for GR and extended gravity approaches. In the 1 + 3 covariant formalism, the perturbations defined describe true physical degrees of freedom and no physical gauge modes exist. In recent years, there has been active research on cosmological perturbations theory for both GR [38–41] and different extended gravity theories [34,35,42] using the 1+3 covariant formalism.

One of the significant advantages of $f(T)$ over $f(R)$ gravity is that its field equations are a second-order instead of fourth-order in the metric. However, $f(T)$ gravity does not respect the local Lorentz invariance, which is one of

a e-mail: shambel.sahlu@wku.edu.et (corresponding author)

the disadvantages of this theory. The teleparallel gravity ($f(T) \equiv T$) is both covariant and local Lorentz invariant [43,44]. But when it comes to the generic function $f(T) \neq T$, the field equations happen to be covariant but not local Lorentz invariant [43]. There is a proposed way to handle this problem. In the work done in [44,45], it has been concluded that in order to achieve local Lorentz invariance for $f(T)$ theory, one has to replace the partial derivative by Lorentz covariant derivative in the definition of the torsion scalar $T^\lambda_{\mu\nu}$ so that one gets a new defined torsion scalar. In the present work, we are not dealing with this type of issue, we rather consider the field equations that were developed in the literature and are covariant [43–45] and the torsion tensor is constructed based on the Weitzenböck connection, focusing on the covariant perturbations [44]. So, we adopted the covariant version of field equations of $f(T)$ gravity from this literature to study the linear cosmological perturbations and depict the significant role of torsion fluid for the growth of density contrasts with redshift using the $1 + 3$ covariant approach. For the case of $f(T) = T$, the field equations reduced to GR which demonstrates that teleparallel gravity and GR are equivalent.

Within the $f(T)$ gravity framework, we will include to the energy-momentum tensor (EMT) of the torsion [fluid] in addition to the EMT of the physical standard matter fluids [44,46] to derive the perturbation equations. After deriving the perturbation evolution equations for generic $f(T)$ theories, we use two paradigmatic power-law $f(T)$ gravity models considered in [47] in $f(T) = \mu T_0 (T/T_0)^n$ and the more generalized $f(T) = T + \mu T_0 (-T/T_0)^n$ as [48]. For certain parametric values, these models produce the accelerating expansion Universe without invoking the cosmological constant. It has been tested to be consistent with the observational data SNIa in [47] and [48] respectively. So, these models are preferred for the cosmological expansion history and now we wish to investigate how they respond to the linear cosmological perturbations in 1 + 3 covariant formalism. For further analysis, we use the well-known approximation technique dubbed quasi-static approximations [13,35,49–51]. For instance, in [13] the validation of such an approximation technique was considered to explore the so-called effective field theory approach to torsional modified gravity by considering the $k^2/a^2 H^2 \gg 1$ regime. In this work, we shall apply this approximation method and assume very slow temporal fluctuations in the perturbations of both the torsion energy density and its momentum compared with the fluctuations of matter-energy density. As such, the time derivative terms of the fluctuations of the torsion field and its momentum are neglected in favor of those matter. Finally, for comparison, we study the growth of energy density fluctuations with redshift for both GR and $f(T)$ gravity approaches.

The road-map of this manuscript is as follows: in the following section, we review the covariant form of the field equations in $f(T)$ gravity. In Sect. 3, the $1 + 3$ covariant gauge-invariant cosmological perturbations formalism within the $f(T)$ gravity framework is presented. The kinematic feature of the Universe and the general fluid description are also studied in the presence of an effective torsion fluid in Sects. 3 and 4 respectively. In Sect. 5, we derive the linear evolution equations for matter and torsion perturbations, the scalar decomposition of which will be carried out in Sect. 6. In Sect. 7, we discuss the harmonic decomposition of the scalar perturbations and figure out how to analyze the growth of matter-energy density perturbations. We explore the growths of matter density contrasts in Sect. 8 for dust and radiation fluids in the GR context and in Sect. 9, for the torsion-dust and torsion-radiation systems for $f(T)$ gravity approach. Finally, we wrap up with the main results of the manuscript in Sect. 10.

## 2 The covariant form of the field equations for $f(T)$ gravity

In this paper, we consider the covariant form of $f(T)$ gravity to clearly show the equivalence between teleparallel gravity, GR and GR + $\Lambda$ as well. This form of field equation is very advisable to define the covariant variables in a gauge-invariant formalism for the study of the cosmological perturbations [44]. To provide the covariant form of the field equation, we start from the relation between both connections, namely: Weitzenböck connection $\tilde{\Gamma}^c_{ab}$ and Levi-Civita connection $\Gamma^c_{ab}$. The torsion tensor is the difference of anti-symmetric part of the Levi-Civita connection [14], and it can be expressed with tetrad fields as[1]

$$T^c_{ab} = e^c_\mu (\partial_a e^\mu_b - \partial_b e^\mu_a).$$ (1)

The difference between the Weitzenböck and Levi-Civita connections

$$K^c_{ab} = \tilde{\Gamma}^c_{ab} - \Gamma^c_{ab},$$ (2)

is called the contortion tensor. The trace of the torsion tensor is given as

$$T = \frac{1}{2} S_d^{ab} T^d_{ab},$$ (3)

where the super-potential term is given as [14,16]

$$S_d^{ab} = K^{ba}_d + \delta^a_d T^{\sigma b}_\sigma - \delta^b_d T^{\sigma a}_\sigma,$$ (4)

and the contortion tensor is re-written as

$$K^{ba}_d = -\frac{1}{2} \left( T^{ab}_d - T^{ba}_d - T_d^{ab} \right).$$ (5)

---

[1] The Latin alphabets represent the tangent space of the manifold, and Greek letters represent for the coordinate on the manifold.

We express the Riemann tensor associated with the Levi-Civita connection and contortion tensors as [43–45]

$$R^d_{cab} = \partial_a \Gamma^d_{cb} - \partial_b \Gamma^d_{ca} + \Gamma^d_{fa}\Gamma^f_{cb} - \Gamma^d_{fb}\Gamma^f_{ca}, \quad (6)$$

$$= \nabla_a K^d_{cb} - \nabla_b K^d_{ca} + K^d_{fa}K^f_{cb} - K^d_{fb}K^f_{ca}, \quad (7)$$

and the Ricci scalar is given as

$$R = -T + 2\nabla^a T_{ab}{}^b \equiv -T + 2\nabla^a T_a, \quad (8)$$

where $T_a = T_{ab}{}^b$. So, the field equations can be rewritten as [45]

$$G_{ab} - \frac{1}{2}g_{ab}T + \nabla^c S_{bca} - S_a^{dc}K_{cdb} = 0, \quad (9)$$

where $G_{ab} = R_{ab} - \frac{1}{2}g_{ab}R$ is well-known Einstein tensor. From this equation, the covariant form of field equations for $f(T)$ gravity yields as [43–45]

$$f'G_{ab} = -\frac{1}{2}g_{ab}\left(f - f'T\right) + \left(f''S_{ab}{}^d \nabla_d T\right) + \kappa^2 \Theta^{(m)}_{ab}, \quad (10)$$

where $g_{ab}$ is the metric, $f' = \mathrm{d}f/\mathrm{d}T$, $f'' = \mathrm{d}^2 f/\mathrm{d}T^2$ and $\Theta^{(m)a}{}_b = \frac{1}{e}\frac{\delta(e\mathcal{L}_m)}{\delta e^b_a}$ denotes the usual EMT of standard matter ($m$) fields.

Note that teleparallel gravity (TG) and GR could be recovered for the limiting case of $f(T) = T$, whereas we restore GR with the cosmological constant $\Lambda$CDM for the case of of $f(T) = T + 2\Lambda$ [52]. It is straightforward to see that the above field equations can be written in the more compact form as

$$G_{ab} = \Theta^{(T)}_{ab} + \Theta^{(m)}_{ab} = \Theta^{eff}_{ab}, \quad (11)$$

where we have defined the EMT of the torsion ($T$) fluid as [12]

$$\Theta^{(T)}_{ab} = -\frac{1}{2f'}g_{ab}(f - f'T)$$
$$-\frac{1}{f'}(f''S_{ab}{}^d \nabla_d T) - \frac{1}{f'}(f' - 1)\Theta^{(m)}_{ab}. \quad (12)$$

All thermodynamic quantities, such as the total energy density $\rho$, isotropic pressure $p$, heat flux $q_a$ and anisotropic stress tensor $\pi_{ab}$ for matter ($m$) and torsion ($T$) fluids are extracted from the total EMT $\Theta_{ab}$ as follows:

$$\rho = \Theta_{ab}u^a u^b, \quad (13)$$

$$p = -\frac{1}{3}h^{ab}\Theta_{ab}, \quad (14)$$

$$q_a = h_a^b u^c \Theta_{bc}, \quad (15)$$

$$\pi_{ab} = h_a^c h_b^d \Theta_{cd} + p h_{ab}, \quad (16)$$

whereas the respective quantities for both matter and torsion components can similarly be extracted from their corresponding EMTs, such that

$$\rho = \rho_T + \bar{\rho}_m, \qquad p = p_T + \bar{p}_m,$$
$$q^a = q_T^a + \bar{q}_m^a, \qquad \pi^{ab} = \pi_T^{ab} + \bar{\pi}_m^{ab},$$

where

$$\bar{\rho}_m = \frac{\rho_m}{f'}, quad \bar{p}_m = \frac{p_m}{f'}, \quad \bar{q}_m^a = \frac{q_m^a}{f'}, \quad \text{and} \quad \bar{\pi}_m^{ab} = \frac{\pi_m^{ab}}{f'}.$$

From Eq. (10), the Friedmann equations of the effective fluid are presented in [12,45] as follows:

$$H^2 = \frac{\rho_m}{3f'} - \frac{1}{6f'}(f - Tf'), \quad (17)$$

$$2\dot{H} + 3H^2 = \frac{p_m}{f'} + \frac{1}{2f'}(f - Tf') + \frac{4f''H\dot{T}}{f'}, \quad (18)$$

where $H(t) \equiv \dot{a}(t)/a(t)$ is the Hubble parameter defined in terms of the scale factor $a(t)$ and its time derivative. One can directly obtain the corresponding thermodynamic quantities such as the effective energy density of the fluid

$$\rho = \frac{\rho_m}{f'} - \frac{1}{2f'}(f - Tf'), \quad (19)$$

and the effective pressure of the fluid

$$p = \frac{p_m}{f'} + \frac{1}{2f'}(f - Tf') + \frac{2f''H\dot{T}}{f'}, \quad (20)$$

respectively. It is easy to show that the Friedmann Eqs. (17) and (18) can be re-expressed as

$$1 = \bar{\Omega}_m + \mathscr{X}, \quad (21)$$

$$\frac{\dot{H}}{H^2} = -\frac{3}{2} + \frac{3w}{2}\bar{\Omega}_m - \frac{3}{2}\mathscr{X} + 3\mathscr{Y}, \quad (22)$$

where we have introduced the following new variables[2]:

$$\mathscr{X} \equiv \frac{Tf' - f}{6H^2 f'}, \ \bar{\Omega}_m \equiv \frac{\rho_m}{3H^2 f'} = \frac{\Omega_m}{f'}, \ \mathscr{Y} \equiv \frac{2\dot{T}f''}{3Hf'}. \quad (23)$$

In this work, we consider the non-interacting perfect fluids and the energy flux and anisotropic stress to be zero in our case. Obviously, in the case of a Lagrangian $f(T) \equiv T$ [43, 44], the physical quantities in Eqs. (12), (19) and (20) reduce to the usual GR limit. In such a limit, the linear cosmological perturbations have been studied in [54,55].

---

[2] It is worth noting here that $\bar{\Omega}_m = \Omega_m/f'$ is the fractional energy density of effective matter like fluid (similar representation is done for $f(R)$ gravity in [34,53], $\Omega_m$ being the normalized energy density parameter of standard matter fluid with $\Omega_m = \Omega_d + \Omega_r$, $\Omega_d$ and $\Omega_r$ being fractional energy densities for dust and radiation and $\mathscr{X}$ being the fractional energy density of torsion fluid alone.

## 3 Kinematic quantities in the presence of torsion

In the $1+3$ covariant decomposition formalism, it is assumed that a fundamental observer slices space-time into temporal and spacial hyper-surfaces [56]. Given the fact that matter components in the Universe would define a physically motivated preferred motion, it is usual to choose the CMB frame, where the radiation dipole vanishes, as the natural reference frame in cosmology [38,57]. For the unperturbed (background) Universe, we define the tangent space-time by the tetrad field $e_0^a = u^a$, where $u^a$ is the four-velocity vector of the observer. The preferred world-line is given in terms of local coordinates $x^a$ in the general coordinate $x^a = x^a(\tau)$ and we define the four-vector velocity $u^a$ as

$$u^a = \frac{dx^a}{d\tau}, \tag{24}$$

where $\tau$ is measured along the fundamental world-line. According to the reason above, the component of any vector $X^a$ parallel to the 4-velocity vector $u^a$ becomes

$$X^a = U_b^a X^b, \quad U_b^a := -u^a u_b, \tag{25}$$

where $U_b^a$ is the projection tensor into the one-dimensional tangent line and satisfies the following relations:

$$U_b^a U_c^b = U_c^a \implies U_b^a u^a = u^a,$$
$$u^a = \delta_0^a \implies U_b^a = \delta_b^a \delta_b^0. \tag{26}$$

Moreover, we define $h_{ab}$ as another projection tensor into the three-dimensional, orthogonal to $u^\mu$ and it satisfies the following properties:

$$h_{ab} = g_{ab} + u_a u_b \implies h_b^a h_c^b = h_c^a,$$
$$h_a^a = 3, \quad h_{ab} u^b = 0. \tag{27}$$

As stated previously, the kinematics of the four-velocity vector $u^a$ determines the geometry of the fluid flow. Any tensor $V_{ab}$ can be expressed as a sum of its symmetric $V_{(ab)}$ and anti-symmetric $V_{[ab]}$ parts as

$$V_{ab} = V_{(ab)} + V_{[ab]}. \tag{28}$$

In this formalism, the covariant derivative of $u_b$ is split into the kinematic quantities [58] as

$$\tilde{\nabla}_a u_b = \frac{1}{3} h_{ab} \tilde{\theta} + \tilde{\sigma}_{ab} + \tilde{\omega}_{ab} - u_a \dot{\tilde{u}}_b, \tag{29}$$

where $\tilde{\theta}$ is the fluid expansion, $\tilde{\sigma}_{ab}$ is the shear tensor, $\dot{\tilde{u}}_a$ is the four-acceleration of the fluid and $\tilde{\omega}_{ab}$ is the vorticity tensor in the presence of torsion. Notice that a tilde represents torsion-dependent physical parameters and a non-tilde represent Levi-Civita connection-dependent parameter. The detailed expressions of torsion dependent kinematic quantities such as expansion of the fluid, shear tensor, the vorticity tensors and the relativistic acceleration vector are presented

in Refs. [17,18,20]. The expansion of the fluid flow in the presence of torsion is given by

$$\tilde{\theta} = \theta - 2u^b T_b, \tag{30}$$

where the torsion vector $T_b$ can be either space-like, time-like or light-like and this three different types of vector torsion is discussed in [17]. Here we have defined the Hubble expansion parameter $3H \equiv \theta$ and $\theta = u^b{}_{;b}$ is the volume-expansion. The shear tensor denotes the change of distortion of the matter flow with time and it is given as

$$\tilde{\sigma}_{ab} = \sigma_{ab} + 2h_a^c h_b^d K_{[cd]}^e u_e, \tag{31}$$

and the vorticity tensor denotes the rotation of matter relative to the non-rotating (Fermi-propagated) frame and it is given as

$$\tilde{\omega}_{ab} = \omega_{ab} + 2h_a^c h_b^d K_{[cd]}^e u_e. \tag{32}$$

Also, the relativistic acceleration vector describes the degree of matter to move under forces other than gravity plus inertia, namely

$$\dot{\tilde{u}}_a = \dot{u}_a + u^b K_{ab}^e u_e, \tag{33}$$

which vanishes for free-falling matter. The general expression for the torsion-based Raychaudhuri equation is given by [17,18,20]

$$\dot{\tilde{\theta}} = \tilde{\nabla}^a \dot{\tilde{u}}_a - \frac{1}{3} \tilde{\theta}^2 - \tilde{\sigma}^{cb} \tilde{\sigma}_{cb} - \tilde{\omega}^{cb} \tilde{\omega}_{cb}$$
$$- R_{cb} u^c u^b - 2u^b T_{cb}^d \left( \frac{1}{3} h_d^c \tilde{\theta} - \tilde{\sigma}_d^c - \tilde{\omega}_d^c - u^c \dot{\tilde{u}}_d \right). \tag{34}$$

In this paper, we assume that the world-line is tangent to $u^c$ but parallel to $\dot{u}_c$, i.e., $u^c \dot{\tilde{u}}_c = 0$. Moreover, $\tilde{\omega}_{cb} = 0 = \tilde{\sigma}_{cb}$ in the case of non-rotational and shear-free fluids and from the covariant approach of the field equation, $R_{cb} u^c u^b = 1/2 (\rho + 3p)$ for relativistic fluid [57,59]. Then, Eq. (34) becomes

$$\dot{\tilde{\theta}} = \tilde{\nabla}^a \dot{\tilde{u}}_a - \frac{1}{3} \tilde{\theta}^2 - \frac{1}{2} (\rho + 3p) - \frac{2}{3} u^b T_b \tilde{\theta}. \tag{35}$$

For a space-like torsion vector the inner product of the torsion and four-velocity vectors of the fluid $u^b T_b$ is vanished identically [20]. Consequently, Eq. (30) reads $\tilde{\theta} = \theta$ and Eq. (33) reads $\dot{\tilde{u}}_a = \dot{u}_a$. Then, from the result of Eq. (35), we obtain

$$\dot{\theta} = -\frac{\theta^2}{3} - \frac{1}{2} (\rho + 3p) + \tilde{\nabla}^a \dot{u}_a, \tag{36}$$

and this equation is the same as the usual Raychaudhuri equation which is presented in Refs. [40,44,57,60].

## 4 General fluid description

Here, we assume the non-interacting matter fluid ($\rho_m \equiv \rho_r + \rho_d$) with torsion fluid in the entire Universe and the growth of the matter-energy density fluctuations has a significant role for formation of large-scale structures.

### 4.1 Matter fluids

Let us consider a homogeneous and isotropic expanding (FLRW) cosmological background and define spatial gradients of gauge-invariant variables such as those of the energy density $D_a^m$ and volume expansion of the fluid $Z_a$ as follows [35,40,61,62]:

$$D_a^m \equiv \frac{a}{\rho_m} \tilde{\nabla}_a \rho_m , \tag{37}$$

$$Z_a \equiv \tilde{\nabla}_a \theta . \tag{38}$$

Those two gradient variables are key to examine the evolution equation for matter density fluctuations.

### 4.2 Torsion fluids

Analogously to the $1 + 3$ cosmological perturbations treatment for $f(R)$ gravity theory [35], let us define extra key variables resulting from spatial gradients of gauge-invariant quantities which are connected with the torsion fluid for $f(T)$ gravity. Accordingly, we define the quantities $\mathscr{F}_a$ and $\mathscr{B}_a$ as

$$\mathscr{F}_a \equiv a \tilde{\nabla}_a T , \tag{39}$$

$$\mathscr{B}_a \equiv a \tilde{\nabla}_a \dot{T} , \tag{40}$$

to characterize the fluctuations in the torsion density and momentum respectively. All the quantities listed in Eqs. (37)–(40) will be considered to develop the system of cosmological perturbation equations for $f(T)$ gravity. Moreover, for each non-interacting fluid, the following conservation equations, considered in [39,40], hold, where

$$\dot{\rho}_m = -\theta(\rho_m + p_m) + (\rho_m + p_m) \tilde{\nabla}^a \Psi_a , \tag{41}$$

and

$$(\rho_m + p_m)\dot{u}_a + \tilde{\nabla}_a p_m + \dot{\Psi}_a - (3c_s^2 - 1)\frac{\theta}{3}\Psi_a + \Pi_a = 0 , \tag{42}$$

hold, where

$$\Psi_a = \frac{q_a}{(\rho_m + p_m)} , \quad \Pi_a = \frac{\tilde{\nabla}^b \pi_{ab}}{\rho_m + p_m} . \tag{43}$$

The speed of sound $c_s^2 = \frac{\delta p}{\delta \rho}$ plays an important role since it allows us to relate the perturbed pressure with the energy density of the fluid. Also, the time derivative of the equation of state parameter $\dot{w} = \dot{p}_m/\dot{\rho}_m$ can be related to the speed of sound [35], and it is given as

$$\dot{w} = (1 + w)(w - c_s^2) . \tag{44}$$

This equation of state parameter is the generalized one for all matter fluids. In fact, for non-interacting fluids, in the following we shall consider the equation of state parameter to be independent of time, thus $\dot{w} = 0$. In this approach, the speed of sound becomes equivalent to the equation of state parameter $w = c_s^2$ [63]. Also, for a perfect fluid both the energy flux and anisotropic-stress are zero ($\Psi_a = \Pi_a = 0$).

## 5 Linear evolution equations

Here we derive the first-order evolution equations for the above-defined gauge-invariant gradient variables. In the energy frame of the matter fluid, these evolution equations for the perturbations are given as:

$$\dot{D}_a^m = -(1 + w)Z_a + w\theta D_a^m , \tag{45}$$

$$\begin{aligned}
\dot{Z}_a &= \left(\frac{w\theta^2}{3(1+w)} - \frac{1+3w}{2f'(1+w)}\rho_m - \frac{w}{2f'(1+w)}(f - Tf') \right. \\
&\quad \left. - \frac{2f''w}{3f'(1+w)}\theta\dot{T} - \frac{w}{1+w}\tilde{\nabla}^2\right)D_a^m \\
&\quad + \left(\frac{2f''}{3f'}\dot{T} - \frac{2\theta}{3}\right)Z_a \\
&\quad - \left(\frac{3\rho_m f''}{2f'^2} + \frac{3w\rho_m f''}{2f'^2} + \frac{2f''^2}{3f'^2}\theta\dot{T} - \frac{2f'''\theta\dot{T}}{3f'}\right)\mathscr{F}_a \\
&\quad + \frac{2f''\theta}{3f'}\mathscr{B}_a , \tag{46}
\end{aligned}$$

$$\dot{\mathscr{F}}_a = \mathscr{B}_a - \frac{w\dot{T}}{1+w}D_a^m , \tag{47}$$

$$\dot{\mathscr{B}}_a = \frac{\dddot{T}}{\dot{T}}\mathscr{F}_a - \frac{w\ddot{T}}{1+w}D_a^m . \tag{48}$$

In the following section, we will see how to decompose the evolution of the above vector gradient variables (45)–(48) into those of scalar variables by applying the scalar decomposition method outlined.

## 6 Scalar decomposition

It is generally understood that the large-scale structure formation follows a spherical clustering mechanism, and that only the scalar (non-solenoidal) parts of the above gradient vectors (45)–(48) assist in the clustering. As a result, we extract the scalar part of a vector $\mathscr{I}_a$ by taking its divergence as [35]

$$a \tilde{\nabla}_a \mathscr{I}_b = \mathscr{I}_{ab} = \frac{1}{3}h_{ab}\mathscr{I} + \Sigma^{\mathscr{I}}_{ab} + \mathscr{I}_{[ab]} , \tag{49}$$

where

$$\mathscr{I} = \tilde{\nabla}_a \mathscr{I}^a, \text{ and } \Sigma_{ab}^{\ \mathscr{I}} = \mathscr{I}_{(ab)} - \frac{1}{3} h_{ab} \mathscr{I}. \tag{50}$$

The last two terms of Eq. (49) describe shear and vorticity effects, respectively. To extract the (scalar) density contrast, the vorticity vanishes and only the shear part is considered. From vector quantities, one can further extract the scalar gradient quantity of our cosmological perturbations, believed to be responsible for the spherical clustering of large-scale structure [35,64]. Let us now define our scalar gradient variables as follows:

$$\Delta_m = a \tilde{\nabla}^a D_a^m, \tag{51}$$

$$Z = a \tilde{\nabla}^a Z_a, \tag{52}$$

$$\mathscr{F} = a \tilde{\nabla}^a \mathscr{F}_a, \tag{53}$$

$$\mathscr{B} = a \tilde{\nabla}^a \mathscr{B}_a. \tag{54}$$

It can be shown that these quantities evolve as:

$$\dot{\Delta}_m = -(1+w)Z + w\theta \Delta_m, \tag{55}$$

$$\dot{Z} = \left[ \frac{w\theta^2}{3(1+w)} - \frac{1+3w}{2f'(1+w)}\rho_m - \frac{w}{2f'(1+w)}(f - Tf') \right.$$
$$\left. - \frac{2f''w}{3f'(1+w)}\theta\dot{T} - \frac{w}{1+w}\tilde{\nabla}^2 \right] \Delta_m + \left[ \frac{2f''}{3f'}\dot{T} - \frac{2\theta}{3} \right] Z$$
$$- \left[ \frac{3\rho_m f''}{2f'^2} + \frac{3w\rho_m f''}{2f'^2} + \frac{2f''^2}{3f'^2}\theta\dot{T} - \frac{2f'''}{3f'}\theta\dot{T} \right] \mathscr{F}$$
$$+ \frac{2f''\theta}{3f'}\mathscr{B}, \tag{56}$$

$$\dot{\mathscr{F}} = \mathscr{B} - \frac{w\dot{T}}{1+w}\Delta_m, \tag{57}$$

$$\dot{\mathscr{B}} = \frac{\ddot{T}}{\dot{T}}\mathscr{F} - \frac{w\ddot{T}}{1+w}\Delta_m. \tag{58}$$

Finally, the second-order scalar evolution equations can be derived by differentiating the above first-order evolution equations with respect to time. For instance, from Eqs. (55) and (56) we obtain

$$\ddot{\Delta}_m = \left[ \frac{1+3w}{2f'}(1-w)\rho_m + \frac{w}{f'}(f - Tf') \right.$$
$$\left. - \frac{2f''w}{3f'}\theta\dot{T} + w\tilde{\nabla}^2 \right] \Delta_m + \left[ \frac{f''}{3f'}\dot{T} + \theta\left(w - \frac{2}{3}\right) \right] \dot{\Delta}_m$$
$$+ \left[ \frac{3\rho_m f''}{2f'^2} + \frac{3w\rho_m f''}{2f'^2} \right.$$
$$\left. + \frac{2f''^2}{3f'^2}\theta\dot{T} - \frac{2f'''}{3f'}\theta\dot{T} \right] (1+w)\mathscr{F}$$
$$- \frac{2f''}{3f'}\theta(1+w)\dot{\mathscr{F}}, \tag{59}$$

whereas from Eqs. (57) and (58) we get

$$\ddot{\mathscr{F}} = \frac{\ddot{T}}{\dot{T}}\mathscr{F} - \frac{2w\ddot{T}}{1+w}\Delta_m - \frac{w\dot{T}}{1+w}\dot{\Delta}_m. \tag{60}$$

The scalar gradient variables (45)–(60) we take as an input to study the energy density fluctuations in different cosmological era by applying the harmonic decomposition of these variables in the next section.

# 7 Harmonic decomposition of variables

From the results of previous sections, we clearly see that the linear cosmological evolution equations of the scalar variables are second-order differential equations, complicated to solve. Thus, in order to obtain the eigenfunctions and corresponding wave-numbers from those second-order differential equations, we shall apply the separation-of-variables technique. Then we shall use the standard harmonic decomposition of the evolution equations for cosmological perturbations [35,61,65] for further details on this technique. All the above linear evolution Eqs. (55)–(60) have a similar structure as the harmonic oscillator equation and the second-order differential evolution equations for any functions $X$ and $Y$ can be represented schematically as [35]

$$\ddot{X} = A\dot{X} + BX - C(Y, \dot{Y}), \tag{61}$$

where the terms $A$, $B$, and $C$ represent the damping oscillator or frictional force, restoring force and source force respectively. Then by applying the separation-of-variables technique, we express

$$X = \sum_k X^k(t)Q^k(\mathbf{x}), \text{ and } Y = \sum_k Y^k(t)Q^k(\mathbf{x}), \tag{62}$$

where $k$ is the wave-number and $Q^k(x)$ is the eigenfunctions of the covariant derivative. Wave-number $k$ represent the order of harmonic oscillator and relate with the scale factor as $k = \frac{2\pi a}{\lambda}$, where $\lambda$ is the wavelength of the perturbations. Here, we define eigenfunctions of the covariant derivative with the Laplace-Beltrami operator for FLRW space-time as

$$\tilde{\nabla}^2 Q^k(x) = -\frac{k^2}{a^2} Q^k(x) . \tag{63}$$

Armed with all this machinery, the first and second-order evolution Eqs. (55)–(60) are expressed as:

$$\dot{\Delta}_m^k = -(1+w)Z_m^k + w\theta \Delta_m^k, \tag{64}$$

$$\dot{Z}^k = \left[ \frac{w\theta^2}{3(1+w)} - \frac{1+3w}{2f'(1+w)}\rho_m \right.$$
$$- \frac{w}{2f'(1+w)}(f - Tf') - \frac{2f''w}{3f'(1+w)}\theta\dot{T} + \frac{wk^2}{a^2(1+w)} \right] \Delta_m^k$$
$$+ \left[ \frac{2f''}{3f'}\dot{T} - \frac{2\theta}{3} \right] Z$$
$$- \left[ (1+w)\frac{3\rho_m f''}{2f'^2} + \frac{2f''^2}{3f'^2}\theta\dot{T} - \frac{2f'''}{3f'}\theta\dot{T} \right] \mathscr{F}^k + \frac{2f''\theta}{3f'}\mathscr{B}^k, \tag{65}$$

$$\dot{\mathscr{F}}^k = \mathscr{B}^k - \frac{w\dot{T}}{1+w}\Delta_m^k, \tag{66}$$

$$\dot{\mathscr{P}}^k = \frac{\ddot{T}}{\dot{T}}\mathscr{P}^k - \frac{w\dot{T}}{1+w}\Delta_m^k, \tag{67}$$

$$\ddot{\Delta}_m^k = \left[\frac{1+3w}{2f'}(1-w)\rho_m + \frac{w}{f'}(f-Tf')\right.$$
$$\left. -\frac{2f''w}{3f'}\theta\dot{T} - \frac{wk^2}{a^3}\right]\Delta_m^k$$
$$+\left[\frac{f''}{3f'}\dot{T} + \theta\left(w-\frac{2}{3}\right)\right]\dot{\Delta}_m^k + \left[(1+w)\frac{3\rho_m f''}{2f'^2}\right.$$
$$+\frac{2f''^2}{3f'^2}\theta\dot{T} - \frac{2f'''}{3f'}\theta\dot{T}\right](1+w)\mathscr{P}^k$$
$$\left. -\frac{2f''}{3f'}\theta(1+w)\dot{\mathscr{P}}^k, \right. \tag{68}$$

$$\ddot{\mathscr{P}}^k = \frac{\ddot{T}}{\dot{T}}\mathscr{P}^k - \frac{2w\ddot{T}}{1+w}\Delta_m^k - \frac{w\dot{T}}{1+w}\dot{\Delta}_m^k. \tag{69}$$

In the following, we shall apply the aforementioned quasi-static approximation in which time fluctuations in the perturbations of the torsion energy density $\mathscr{P}^k$ and momentum $\mathscr{B}^k$ are assumed to be constant with time, i.e., one is allowed to take $\ddot{\mathscr{P}}^k = \ddot{\mathscr{B}}^k = \dot{\mathscr{B}}^k \approx 0$. Under this approximation, the first-order linear evolution Eqs. (64)–(65) reduce to:

$$\dot{\Delta}_m^k = -(1+w)Z_m^k + w\theta\Delta_m^k, \tag{70}$$

$$\dot{Z}^k = \left[\frac{w\theta^2}{3(1+w)} - \frac{1+3w}{2f'(1+w)}\rho_m - \frac{w}{2f'(1+w)}(f-Tf')\right.$$
$$+\frac{wk^2}{a^2(1+w)}\right]\Delta_m^k + \left(\frac{2f''}{3f'}\dot{T} - \frac{2\theta}{3}\right)Z$$
$$-\left(\frac{3\rho_m f''}{2f'^2} + \frac{3w\rho_m f''}{2f'^2}\right.$$
$$\left. +\frac{2f''^2}{3f'^2}\theta\dot{T} - \frac{2f'''}{3f'}\theta\dot{T}\right)\mathscr{P}^k. \tag{71}$$

Also, from Eqs. (66) and (69) results the relation

$$\mathscr{P}^k = \frac{2w\dot{T}\ddot{T}}{(1+w)\dddot{T}}\Delta_m^k + \frac{w\dot{T}^2}{(1+w)\dddot{T}}\dot{\Delta}_m^k. \tag{72}$$

By using the Eq. (72) together with the quasi-static approximation itself, Eq. (68) for matter-energy density perturbations yields

$$\ddot{\Delta}_m^k = \left\{\frac{1+3w}{2f'}(1-w)\rho_m + \frac{w}{f'}(f-Tf')\right.$$
$$-\frac{2f''w}{3f'}\theta\dot{T} - w\frac{k^2}{a^2}$$
$$+(1+w)\left[\frac{3\rho_m f''}{2f'^2} + \frac{2f''^2}{3f'^2}\theta\dot{T} - \frac{2f'''}{3f'}\theta\dot{T}\right]\frac{2w\dot{T}\ddot{T}}{\dddot{T}}\right\}\Delta_m^k$$
$$+\left[\frac{f''}{3f'}\dot{T} + \theta\left(w-\frac{2}{3}\right)\right.$$
$$\left. +\left(\frac{3\rho_m f''}{2f'^2} + \frac{3w\rho_m f''}{2f'^2} + \frac{2f''^2}{3f'^2}\theta\dot{T} - \frac{2f'''}{3f'}\theta\dot{T}\right)\frac{w\dot{T}^2}{\dddot{T}}\right]\dot{\Delta}_m^k. \tag{73}$$

For the case of $f(T) = T + 2\Lambda$, Eq. (73) is reduced to the well-known evolution equation of $\Lambda CDM$:

$$\ddot{\Delta}_m^k = \left[\frac{3}{2}\Omega_m(1+3w)(1-w)H^2 + 6wH^2\Omega_\Lambda - w\frac{k^2}{a^2}\right]\Delta_m^k$$
$$+3H\left(w-\frac{2}{3}\right)\dot{\Delta}_m^k, \tag{74}$$

where the energy density of cosmological constant fluid $\Lambda = 3H_0^2\Omega_\Lambda$. In this context, the matter and cosmological constant fluids are involved in the growth of the energy density fluctuations. Also, for the paradigmatic case of $f(T) = T$ [43,44], GR is exactly recovered and the evolution Eq. (73) coincides with GR as [34]

$$\ddot{\Delta}_m^k = \left[\frac{3}{2}\Omega_m(1+3w)(1-w)H^2 - w\frac{k^2}{a^2}\right]\Delta_m^k$$
$$+3H\left(w-\frac{2}{3}\right)\dot{\Delta}_m^k. \tag{75}$$

As we shall see in the following sections, Eq. (73) remains a key equation for analyzing the growth of energy density fluctuations capable of explaining the formation of large-scale structures. For the sake of simplicity and with the aim of illustrating the versatility of our analysis, we shall consider a paradigmatic power-law $f(T)$ gravity models which is compatible with the cosmic acceleration for $n > 1.5$ [47], and the model is given as

$$f(T) = \mu T_0 \left(\frac{T}{T_0}\right)^n, \tag{76}$$

where $\mu$ and $n$ are dimensionless constants, and $T_0 = -6H_0^2$ is the present-day value of the torsion scalar. For the case of $n = 1$ this model is suitable to recover GR. Let us further assume a power-law expansion [66–68]:

$$a(t) = a_0(t/t_0)^m, \tag{77}$$

where $m$ is a positive constant, and as usual the scale factor[3] is related to the cosmological redshift as $a = a_0/(1+z)$.

From Eq. (21), we redefine the normalized energy density parameter for non-interacting torsion-matter fluids as $1 = \bar{\Omega}_d + \Omega_T$, since, $\Omega_T \equiv \mathscr{X}$ is the normalized energy density parameter of torsion fluid. Consequently, a normalize parameters for the fluid yields

$$\bar{\Omega}_m = \frac{2n-1}{n}. \tag{78}$$

With this definition, it is possible to know the amount of matter fluid in the non-interacting system and analyze the growth of matter fluctuations with redshift (we will see in detail for torsion-dust and torsion-radiation cases in Sect. 9)[4]. As an example: for $n = 1$, the matter fluid is large enough in the system and torsion fluid becomes negligible. In this case, we obtain the matter-dominated Universe and our generalized evolution Eq. (73) reduces to Eq. (75). For the case of $n \approx 0.595$, the value of $\tilde{\Omega}_m = \Omega_m \approx 0.32$ [69], consequently $\Omega_T = \Omega_\Lambda \approx 0.68$ [69], with the understanding that the torsion fluid acts as a cosmological constant. This indicates that our system filled by both fluids: matter and torsion fluid. For $n \geq 1$, $\tilde{\Omega}_m \geq 1$ and $\Omega_T \leq 0$, in this situation the matter fluid is a major component of the Universe and contributions of torsion like a fluid with a negative energy density are the same as of a cosmological constant. However, the effective energy density of the fluid becomes $\rho = \rho_m + \rho_T \geq 0$ [12] as presented in Eq. (19). Before solving the linear evolution Eqs. (70)–(73), let us point out that for $f(T)$ gravity model (76) and scale factor (77), the background quantities $\Omega_T$ and $\mathscr{Y}$ as defined in Eq. (23) become

$$\mathscr{Y} = \frac{(n-1)\Big[2w(2n-1)-1\Big]}{n(5-4n)}, \tag{79}$$

$$\Omega_T = \frac{1-n}{n}, \quad n \neq 0. \tag{80}$$

---

[3] In this manuscript, both $a_0$ and $t_0$ are normalized to unity for simplicity.

For the case for $n = 1$, $\mathscr{Y} = \Omega_T = 0$. Here, we define the normalized energy density contrast for matter fluid as

$$\delta^k(z) \equiv \frac{\Delta_m^k(z)}{\Delta_{in}}, \tag{81}$$

where the subscript $in$ refers initial value of $\Delta_m(z)$ at redshift $z_{in}$.[5] Where analogously to (81)

$$\delta_{GR} \equiv \frac{\Delta_m^k(z)(n=1)}{\Delta_i(z_{in})}. \tag{82}$$

and for the case for $f(T) = T$, $\delta^k(z) = \delta^k_{GR}(z)$ which coincides with TEGR and the results are exactly the same as GR. Indeed, the variation of CMB temperature detected observationally in the order of $10^{-5}$ [70] and this variation strongly supports the gravitational perturbations initially through their redshifting effect on the CMB [71,72]. Also, we shall assume the following initial conditions as $\Delta_{in} \equiv \Delta^k(z_{in} = 1100) = 10^{-5}$ and $\dot{\Delta}_{in} \equiv \dot{\Delta}^k(z_{in} = 1100) = 0$, for every mode $k$ to deal with the growth of matter fluctuations (similar analysis is done in [53]). Therefore the energy density fluctuations have initial value as $\Delta_m(z_{in}) = \Delta_{in} = 10^{-5}$ for all $n$ at the initial redshift $z_{in} = 1100$. At this redshift the value of normalized energy density perturbations of the matter fluid presented in Eq. (81) becomes one ($\delta(z_{in}) = 1$).

For convenience, we also transform any time derivative functions $f$ and $H$ into a redshift derivative as follows:

$$\frac{\dot{f}}{H} = \frac{df}{dN}, \quad \text{where} \quad N \equiv \ln(a), \tag{83}$$

$$\dot{f} = -(1+z)H\frac{df}{dz}, \quad \text{and}$$

$$\ddot{f} = (1+z)^2 H\left(\frac{dH}{dz}\frac{df}{dz} + H\frac{d^2f}{dz^2}\right). \tag{84}$$

We apply these transformation techniques in the following sections.

## 8 Matter density fluctuations in GR and $\Lambda$CDM Limits

In this section, we analyze the growth of energy density fluctuations for dust and radiation fluids in $\Lambda$CDM and GR limits from Eqs. (74) and (75) respectively.

### 8.1 Dust-dominated universe

If, we assume that the Universe is dominated by dust fluid only, then the equation of state parameter is $w_d \approx 0$. Consequently, Eqs. (74) and (75) read [6]

$$\ddot{\Delta}_d + 2H\dot{\Delta}_d - \frac{3}{2}\Omega_d H^2 \Delta_d = 0. \tag{85}$$

By applying Eqs. (83) and (84), this equation becomes

$$\frac{d^2\Delta_d(z)}{dz^2} - \frac{1}{2(1+z)}\frac{d\Delta_d(z)}{dz} - \frac{3\Omega_d\Delta_d(z)}{2(1+z)^2} = 0. \tag{86}$$

and admits the solution

$$\Delta(z) = C_1(1+z)^{\frac{1}{4}(1+\sqrt{24\Omega_d+9})} + C_2(1+z)^{\frac{1}{4}(1-\sqrt{24\Omega_d+9})}, \tag{87}$$

where $C_1$ and $C_2$ are integration constants and we determine these constant by imposing the above initial conditions[7]. In the dust-dominated Universe, the input parameter $\Omega_d$ is a key point to determine the magnitude of matter fluctuations with redshift. For instance, the numerical result of Eq. (87) is presented in Fig. 1 for $\Omega_d = 1$ (blue-solid) and it shows the growth of energy density fluctuations of dust fluid in the dust-dominated Universe. With this plot, the density contrasts grows up with redshift.

---

[4] For the case of $n < 0.5$, the normalized effective fluid energy density parameter has a negative sign which shows an unphysical mode based on relation (78).

[5] We set the initial conditions at $z_{in} \approx 1100$ during the decoupling era. So, in the following two sections we shall explore the feature of fractional energy density perturbations $\delta(z)$ with redshift $0 \leqq z \leqq 1100$.

[6] From here onwards, we remove the superscript $k$ to avoid the unnecessary cluttering of notations.

[7] **NB** All integral constants $C_i$, $i = 1, 2, \ldots 18$ are determined by imposing the initial conditions $\Delta_{in} = 10^{-5}$ and $\dot{\Delta}_{in} = 0$. And we used these initial conditions to present all plots in this manuscript.

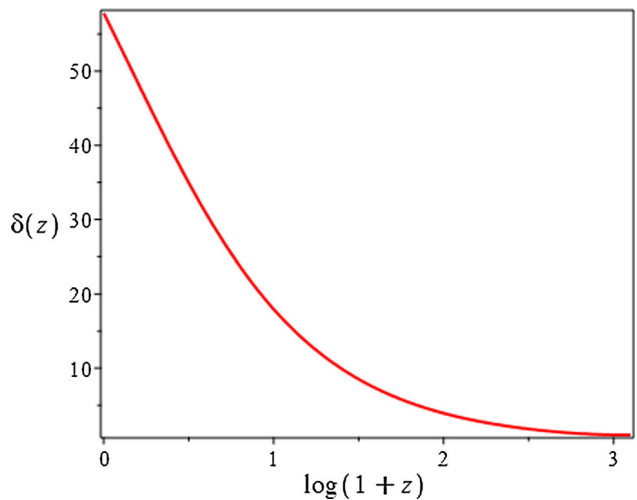

**Fig. 1** The $\delta(z)$ versus redshift $z$ for Eq. (87) for dust dominated Universe for GR limit

## 8.2 Radiation-dominated universe

Here, we study the growth of energy density perturbations of the radiation fluid, by assuming the Universe has two non-interacting cosmic fluid components, namely radiation and the cosmological constant. In this assumption, the equation of state parameter $w_r \approx 1/3$ and the normalized energy density parameters is $\Omega_\Lambda = 1 - \Omega_r$. Then, Eq. (74) in redshift space becomes

$$\frac{d^2 \Delta_r(z)}{dz^2} + \frac{1}{1+z}\frac{d\Delta_r(z)}{dz} - \frac{1}{(1+z)^2}\left[2(\Omega_r + \Omega_\Lambda) - \frac{k^2}{3a^2 H^2}\right]\Delta_r(z) = 0. \quad (88)$$

Notice that, the difference between $\Lambda$CDM and GR limits is the parameter $\Omega_\Lambda$ in sence that $\Omega_\Lambda = 0$ in GR solutions but remains in the $\Lambda$CDM. So, for the radiation-dominated Universe, we can re-write Eq. (88) as

$$\frac{d^2 \Delta_r(z)}{dz^2} + \left(\frac{1}{1+z}\right)\frac{d\Delta_r(z)}{dz} - \frac{1}{(1+z)^2}\left[2\Omega_r - \frac{16\pi^2}{3\lambda^2(1+z)^4}\right]\Delta_r(z) = 0 . \quad (89)$$

For $\Lambda$CDM limits $\Omega_r + \Omega_\Lambda = 1$ and for GR limit $\Omega_\Lambda = 0$, consequently the solution of Eqs. (88) and (89) have the same behavior, and choose Eq. (88) to present the numerical solutions.

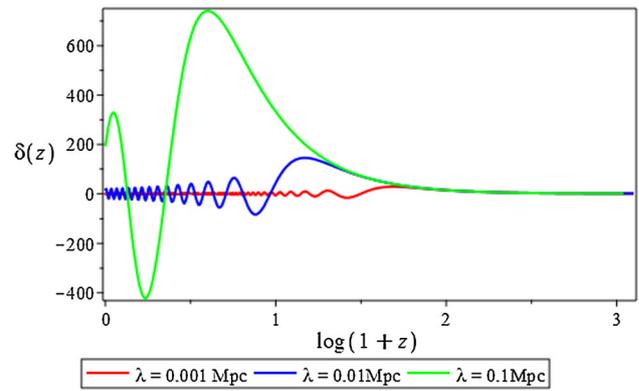

**Fig. 2** $\delta(z)$ versus $z$ for Eq. (90) for short-wavelength mode in the radiation-dominated Universe for $\Lambda$CDM approach. We consider different wave-length ($\lambda$) limits and $1 = \Omega_r + \Omega_\Lambda$ for plotting

The exact solution of Eq. (88) for the short-wavelength mode, $k^2/a^2 H^2 \gg 1$, becomes

$$\Delta_r(z) = C_3 \text{BesselJ}\left(\frac{1}{2}\sqrt{2(\bar{\Omega}_r + \Omega_\Lambda)}, \frac{2}{3}\frac{\sqrt{3}\pi}{\lambda}\frac{1}{(1+z)^2}\right)$$
$$+ C_4 \text{BesselY}\left(\frac{1}{2}\sqrt{2(\bar{\Omega}_r + \Omega_\Lambda)}, \frac{2}{3}\frac{\sqrt{3}\pi}{\lambda}\frac{1}{(1+z)^2}\right), \quad (90)$$

where $C_3$ and $C_4$ are the are integration constants. Whereas in the long-wavelength mode, $k^2/a^2 H^2 \ll 1$, the exact solution reads

$$\Delta(z) = \log(1+z)\left[C_5 \sinh\left(\sqrt{2(\Omega_r + \Omega_\Lambda)}\right) + C_6 \cosh\left(\sqrt{2(\Omega_r + \Omega_\Lambda)}\right)\right]. \quad (91)$$

In the above result, the relation $\frac{k^2}{a^2 H^2} = \frac{16\pi^2}{\lambda^2(1+z)^4}$ is used. The numerical results of the matter density contrast for Eqs. (90) and (91) are presented in Figs. 2 and 3, respectively. From these plots, we see that the energy density fluctuations of a radiation fluid are growing on large wave-length scales with decrease in redshift (see Fig. 3) but oscillate in the short wave-length limits for different values of $\lambda$ (see Fig. 2)[8].

## 9 Matter density fluctuations in $f(T)$ gravity approach

Here, we consider the cosmic medium as a mixture of two non-interacting fluids as a torsion-dust and torsion-radiation mixture.

---

[8] We consider small value of $\lambda = 0.001, 0.01$ and $0.1$ in Mpc for short-wave length cases.

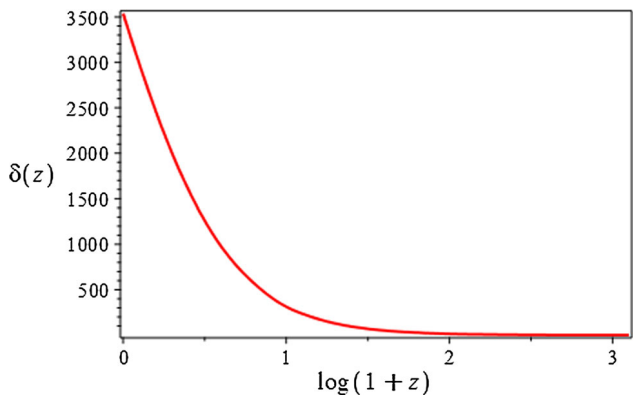

**Fig. 3** $\delta(z)$ versus $z$ for Eq. (91) for long-wavelength mode in radiation-dominated Universe for $\Lambda$CDM approach and for $1 = \Omega_r + \Omega_\Lambda$

### 9.1 Torsion-dust system

In this fluid mixtures, we assume that the Universe hosts two dominant cosmological fluids, namely a torsion-like fluid and the usual dust ($w_d = 0$) matter. In this case, evolution Eq. (73) reduces to

$$\ddot{\Delta}_d = \frac{\rho_d}{2f'}\Delta_d + \left(\frac{2f''}{3f'}\dot{T} - \theta\frac{2}{3}\right)\dot{\Delta}_d, \tag{92}$$

since $\Delta_m \approx \Delta_d$, $\rho_m/f' = \rho_d/f'$, consequently, $\bar{\Omega}_m = \bar{\Omega}_d$ is the effective normalized dust density.

As we compare this perturbation equation with the equations for perturbations in the metric approach as presented in [30], the extra term $\frac{2f''\dot{T}}{3f'} = H\mathscr{Y}$ is obtained due to the definition of the spatial gradients of gauge-invariant variables, Consequently, the results of the matter density contrasts have different features for both approaches. For the assumptions of $\mathscr{Y} = 0$ (or if the rate of change of the torsion scalar is very slow, i.e., $\dot{T} \approx 0$), our result is mathematically the same as [30] in the dust-torsion system.

Then, we choose our paradigmatic $f(T)$ gravity model from Eq. (76) and applying Eq. (79) onto (108) yields[9]

$$\ddot{\Delta}_d - H(\mathscr{Y} - 2)\dot{\Delta}_d - \frac{3}{2}H^2\bar{\Omega}_d\Delta_d = 0. \tag{93}$$

We note that for the case of $n = 1$, the parameter $\mathscr{Y} = 0$ and Eq. (97) reduces to the well-known evolution equation of energy density of dust fluid in the GR limit namely

$$\ddot{\Delta}_d + 2H\dot{\Delta}_d - \frac{3}{2}H^2\Omega_d\Delta_d = 0. \tag{94}$$

---

[9] $\mu$ is eliminated during the simplification.

In redshift space, it can be shown that Eq. (93) yields

$$\frac{d^2\Delta_d}{dz^2} + \frac{1}{1+z}\left(\mathscr{Y} + \frac{1}{m} - 2\right)\frac{d\Delta_d}{dz} - \frac{3\bar{\Omega}_d\Delta_d}{2(1+z)^2} = 0. \tag{95}$$

Our free parameters $\bar{\Omega}_d$, $m$ and $n$ have a significant role to play in the numerical solution of Eq. (95), and explore the fluctuation of energy density with redshift. To provide the parameter $\bar{\Omega}_d$, we use the definition from Eq. (78). From this definition, it is possible to determine the fractional amount of the normalized energy density parameters $\Omega_T$ and $\bar{\Omega}_d$ in the system. For instance, at $n = 1$, $\bar{\Omega}_d = 1$ and $\Omega_T$ reads zero. In this case, the numerical solution reduces to GR limit Eq. (86) and the effective matter fluid acts as dust. For $n \geq 1$, $\bar{\Omega}_d \geq 1$ and $\Omega_T \leq 0$, the dust fluid is the major component of the Universe and we note the contributions of torsion fluid with negative energy density but $\rho = \rho_d + \rho_T \geq 0$. For $n = 0.9$, $\bar{\Omega}_d \approx 0.88$ and $\Omega_T \approx 0.12$ the Universe has relatively more dust than torsion fluids and at a particular $n \approx 0.5953$, the value of normalized energy density parameter $\bar{\Omega}_d \approx 0.32$ as the observation expects $\Omega_d \approx 0.32$ in SNIa data. Consequently, $\Omega_T = 0.6800001$ which closer to the observed value of $\Omega_\Lambda = 0.68$ and the torsion fluid acts as cosmological constant. At $n \approx 0.595$, the growth of energy density fluctuations occurs in the present torsion-dust era. For the case of $n = 0.5$, $\bar{\Omega}_d$ reads zero and $\Omega_T = 1$, i.e., the Universe becomes torsion dominated fluid alone at the background level.

Due to cosmic expansion, the background energy density of the dust fluid decreases with the scale factor of Universe, $\rho = \rho_0 a^{-3}$ and it is proportional to the redshift $z$. Then, the scale factor becomes $a(t) = a_0(t/t_0)^{2/3(1+w_d)}$. To keep the generality our anastz in Eq. (77), we choose $m = 2/3(1 + w_d) = 2/3$ for the scale factor exponent in Eq. (77) assuming a dust-dominated epoch. Then, we substitute Eqs. (78) and (79) into our evolution equation which (95), consequently reads

$$\frac{d^2\Delta_d}{dz^2} + \frac{1}{1+z}\left(\frac{(1-n)}{n(5-4n)} - \frac{1}{2}\right)\frac{d\Delta_d}{dz} - \frac{3(2n-1)}{2n(1+z)^2}\Delta_d = 0. \tag{96}$$

This equation is similar to the Euler–Cauchy type equation in $z$ which is presented in [73] and it can be rewritten as

$$(1+z)^2\frac{d^2\Delta_d}{dz^2} + (1+z)b\frac{d\Delta_d}{dz} - c\Delta_d = 0, \quad \text{since}$$
$$b = \frac{1-n}{n(5-4n)} - \frac{1}{2}, \quad c = \frac{3(2n-1)}{2n}. \tag{97}$$

Let us consider $\Delta(z) = (1+z)^r$, and the first- and second-order derivative with respect to redshift becomes $r(1+z)^{r-1}$ and $r(r-1)(1+z)^{r-1}$ respectively. Then, the characteristic

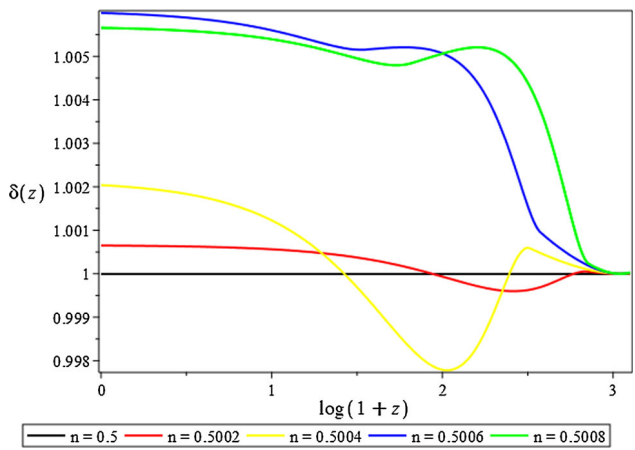

**Fig. 4** $\delta(z)$ versus $z$ for Eq. (100) in the torsion-dust system for $n$ is closer to 0.5

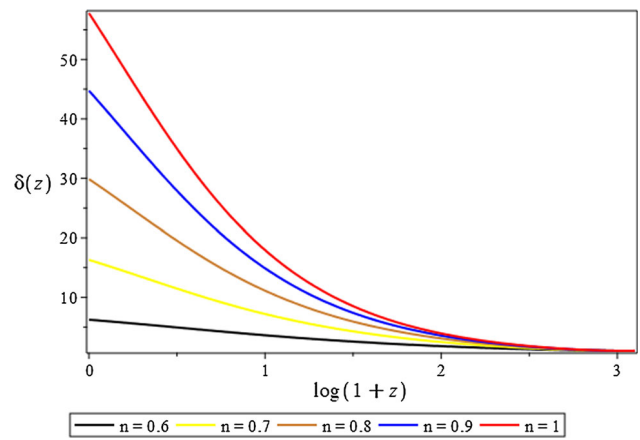

**Fig. 5** $\delta(z)$ versus $z$ for Eq. (100) in the torsion-dust system for $n$ closer to 1

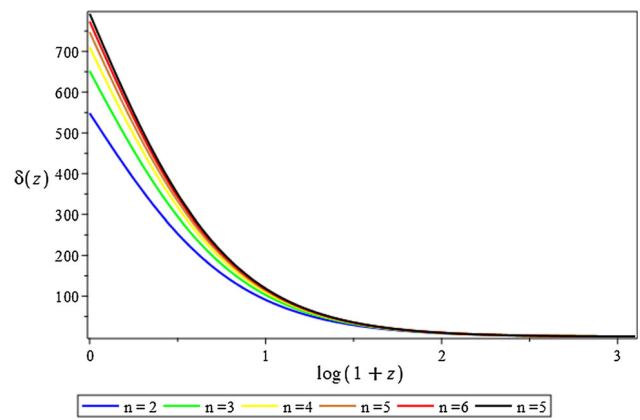

**Fig. 6** $\delta(z)$ versus $z$ for Eq. (100) in the dust-torsion system for $n > 1.5$

equation reads

$$r^2 + r(b - 1) - c = 0, \tag{98}$$

and the solution is given by

$$\begin{aligned} r_\pm &= \frac{-b + 1 \pm \sqrt{b^2 - 2b + 4c + 1}}{2} \\ &= \frac{(12n^2 - 17n + 2 \pm \varpi)}{4n(4n - 5)}. \end{aligned} \tag{99}$$

The value of $r_\pm$ is always real for $n \geq 0.5$. Then, the exact solution of Eq. (97) is given as

$$\begin{aligned} \Delta_d(z) &= C_7(1 + z)^{\frac{(12n^2 - 17n + 2 + \varpi)}{4n(4n - 5)}} \\ &\quad + C_8(1 + z)^{\frac{(12n^2 - 17n + 2 - \varpi)}{4n(4n - 5)}}, \end{aligned} \tag{100}$$

where $\varpi = \sqrt{912n^4 - 274n^3 + 2497n^2 - 668n + 4}$. Our free parameter $n$ has a significant role to present the numerical solution of Eq. (100), and to explore the growth of energy density fluctuations with redshift. However, for illustrative purposes, we use different values of $n$ for all numerical analysis and show that the $f(T)$ gravity model under consideration is an alternative approach to explain the growth of matter fluctuations in a dust-dominated Universe and make a comparison with the well-known theory of gravity GR limit as well.[10]

Now, we only use one free parameter $n$ to present the numerical results of the torsion-dust system Eq. (100). We choose different ranges of $n$ and clearly see the behavior of the growth of density contrasts with cosmic-time. For example, at $n = 0.5$, $\Omega_m = 0$ (black line) and $\Omega_T = 1$ in Fig. 4, and observe that the rate of the growth energy density fluc-tuations is constant[11] with cosmic-time ($\delta(z) = 1$), but it is growing with redshift till near-future epoch for $n > 0.5$. The growth of density contrasts (in this case the dust density contrast) is decreasing instead of for $n < 0.5$ which is unrealistic for growth of perturbations and is growing with cosmic time for $n \geq 0.5$. Based on these facts, we use $n \geq 0.5$. The growth of density constants has a near-oscillatory behavior for $n$ closer to 0.5. On-the other hand, for the case of $n = 1$ (blue solid line) in Fig. 5, the growth of density fluctuations is the same as GR which is presented in Fig. 1. The growth of the fluctuations also is proportional to the values of $n$. For instance, at $n > 1$ in Fig. 6, the growth of energy density fluctuations is very high compared with other values of $n$ presented in the other plots. As mentioned earlier, for $n$ closer to 0.5 the second term of RHS in Eq. (100) is over-dominated than the first term and due to this reason, the behavior of $\delta(z)$ shows small oscillation for the intervals of

---

[10] The first terms of the right-hand side of Eq. (100) is growing with redshift so decaying with cosmic-time, and the second term is decayed with redshift so growing with cosmic-time for any value of $n$.

[11] If $\delta(z) = 0$, density fluctuations is nil with time; for $\delta(z) = constant$, mean that the rate of growth density fluctuations is constant with time.

**Table 1** We illustrate the features of density contrast for torsion-dust system with different ranges of $n$

| Ranges of $n$ | Behavior of $\delta(z)$ |
| --- | --- |
| $n = 0.5$ | Growing with a constant rate |
| $0.5 < n \leq 0.5008$ | Oscillating |
| $n > 0.5008$ | Growing |

We consider here $n \geq 0.5$ due to the reason of definition (78)

$0.5 < n < 0.5008$. However, for $n$ greater than 0.5008, the energy densities contrast $\delta(z)$ is growing with cosmic time and it is also growing proportional to $n$ values. Moreover, we discuss the behavior of the growth of density contrasts in Table (see Table 1).

The growth of the matter density perturbations can be explored for the range of $n \geq 0.5$. In deed, the work done by Wei et.al in [47] pointed-out that the value of $n > 1.5$, for the deceleration parameter $q$ to be negative and to explain the background history of the accelerating expansion. Within this frame-work, the deceleration parameter can be given by

$$q = -1 - \frac{\dot{H}}{H^2} = -1 + \frac{3}{2} + \frac{3w_d}{2}\bar{\Omega}_d + \frac{3}{2}\mathscr{X} - 3\mathscr{Y}$$
$$= -1 + \frac{3}{2} + \frac{3}{2}\mathscr{X} - 3\mathscr{Y}; \quad \begin{cases} q < 0, & \text{for } n > 1.5, \\ q \geq 0, & \text{for } n \leq 1.5. \end{cases}$$
$$(101)$$

Then, $q$ has positive values from $0.5 \leq n \leq 1.5$, and negative values for $n > 1.5$. So, we can easily identify the growth of density contrasts in the decelerating and accelerating epochs by using the above relations of $n$ and $q$. We note here that the range of $n$ between 0.5 to 1.5 is incompatible with current observations (of an accelerated expansion phase). To solve such inconsistency between model and observation, we look at the more general $f(T)$ gravity model

$$f(T) = T + \mu T_0 \left( -\frac{T}{T_0} \right)^n, \quad (102)$$

which is constrained from solar-system tests to be valid only for small values of $n \ll 1$ as [48]. $\Lambda$CDM can be recovered for the case of $n = 0$. From the Eq. (19), the parameter $\mu$ yields as

$$\mu = \left( 6H_0^2 \right)^{1-n} \left( \frac{1 - \Omega_{(m=d)}}{2n - 1} \right). \quad (103)$$

By applying the same mathematical procedures, the exact solution of Eq. (108) for this model reads as

$$\Delta_d(z) = C_9 (1+z)^{l_+} + C_{10} (1+z)^{l_-}, \quad (104)$$

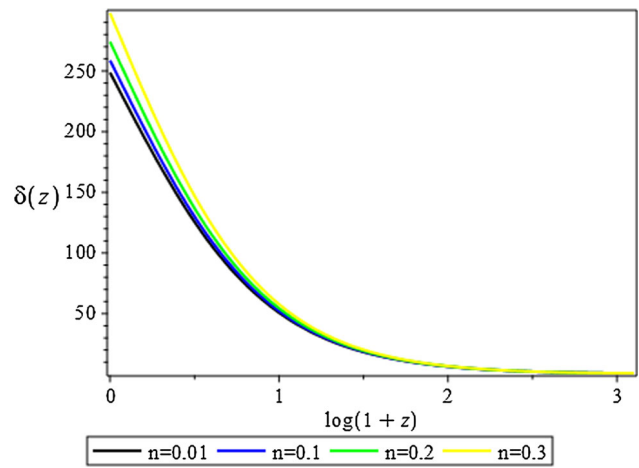

**Fig. 7** $\delta(z)$ versus $z$ for $f(T) = T + \mu T_0 (T/T_0)^n$ in the torsion-dust system for for different values of $n$. We used $\Omega_d = 0.32$ for illustrate purposes

where

$$l_\pm = \frac{\mathscr{Y}_1}{4} + \frac{1}{2} \pm \frac{1}{4}\sqrt{\mathscr{Y}_1^2 + 4\mathscr{Y}_1 - 24\Omega_T + 28} \quad (105)$$

$$\Omega_T = \frac{(1+n)(1 - \Omega_d)}{1 - 2n + n(1 - \Omega_d)} \quad (106)$$

$$\mathscr{Y}_1 = \frac{12n(n-1)\Omega_d(\Omega_d + n + 1)}{(n\Omega_d + n - 1)(23n\Omega_d - 24\Omega_d - n + 1)}. \quad (107)$$

Note that, for the case of $n = 0$, we have $\Omega_T = 1 - \Omega_d = \Omega_\Lambda$ and $\mathscr{Y}_1 = 0$. The numerical results of Eq. (104) are presented in Fig. 7. From this figure, we clearly show that the matter density contrast is growing with cosmic-time.

This shows that the $f(T)$ gravity model Eq. (102) is viable to study the matter density contrast, formation of large-scale, and it is also favored with the observational bounds. And the amplitude of matter density contrasts is increasing in the torsion-dust Universe for $n \ll 1$ which is favored with the theoretical and observational aspects of cosmology.

### 9.2 Torsion-radiation system

Here, we assume that the Universe is dominated by a torsion fluid and radiation ($w_r = 1/3$) mixture as a background, consequently the energy density of dust matter contribution is negligible. In such a system, perturbations would evolve according to the following Eq. (73)

$$\ddot{\Delta}_r = \left[ \frac{2}{3f'}\rho_r + \frac{1}{3f'}(f - Tf') - \frac{2f''}{9f'}\theta\dot{T} \right.$$
$$- \frac{k^2}{3a^2} + \left( \frac{2\rho_r f''}{f'^2} + \frac{2f''^2}{3f'^2}\theta\dot{T} \right.$$
$$\left. - \frac{2f'''}{3f'}\theta\dot{T} \right) \frac{2\dot{T}\ddot{T}}{3\ddot{T}} \right]\Delta_r + \left[ \frac{f''}{3f'}\dot{T} \right.$$

$$+\frac{\theta}{3}+\left(\frac{2\rho_r f''}{f'^2}+\frac{2f''^2}{3f'^2}\theta\dot{T}-\frac{2f'''}{3f'}\theta\dot{T}\right)\frac{\dot{T}^2}{3\dot{T}}\right]\dot{\Delta}_r,$$
$$(108)$$

i.e., $\Delta_m \approx \Delta_r$, $\rho_m = \rho_r$ and $\bar{\Omega}_m = \Omega_r$. By applying our paradigmatic $f(T)$ gravity model and the power scale factor associated with Eqs. (79) and (80), it can be shown that the second-order evolution Eq. (108) of the energy density for torsion-radiation system can be re-written as

$$\ddot{\Delta}_r - H\left(\frac{\mathscr{Y}}{2}\left\{1+\frac{m}{3}\left[\bar{\Omega}_r(1-n)-2n+3\right]\right\}-1\right)\dot{\Delta}_r$$
$$-H^2\left[2n\bar{\Omega}_r - 2\Omega_T - \mathscr{Y}(6-2n) - \frac{k^2}{3H^2a^2}\right]\Delta_r = 0.$$
$$(109)$$

For $n = 1$, this equation reduces to the well-known GR limit [34]:

$$\ddot{\Delta}_r + H\dot{\Delta}_r - \left(2H^2\bar{\Omega}_r - \frac{k^2}{3a^2}\right)\Delta_r = 0. \qquad (110)$$

In the following two sub-sections, we further analyze the growth of energy density fluctuations from the evolution Eq. (109) in short- and long-wavelength modes.

### 9.2.1 Short-wavelength mode

Here, we discuss the growth of fractional energy density fluctuations within the horizon, where $k^2/a^2H^2 \gg 1$. In this regime, the Jeans wavelength $\lambda_J$ is much larger than the wavelength of the mean free path of the photon $\lambda_p$ and the wavelength of the non-interacting fluid, i.e., $\lambda \ll \lambda_p \ll \lambda_J$ (see similar analysis: [39] for GR and [35] for $f(R)$ gravity theory approaches).

For further processing, we have to use the definitions of (78), (79), and apply the same reason to fix the first parameter $m$ as Sect. 9.1 for expanding Universe $\rho = \rho_0 a^{-4}$, and the scale factor becomes $a(t) = a_0(t/t_0)^{2/3(1+w_r)}$. Explicitly, we can choose our input parameter $m = 2/3(1 + w) = 1/2$ for the scale factor exponent in Eq. (77) assuming we are in a radiation-dominated epoch. In this context our leading Eq. (109) reads

$$\frac{d^2\Delta_r}{dz^2} + \frac{\beta}{1+z}\frac{d\Delta_r}{dz}$$
$$-\frac{1}{(1+z)^2}\left[\gamma - \frac{16\pi^2}{3\lambda^2(1+z)^4}\right]\Delta_r(z) = 0. \qquad (111)$$

where

$$\beta \equiv \frac{4n^4 + 33n^3 - 57n + 2}{72n^3 - 90n^2},$$
$$\text{and} \quad \gamma \equiv \frac{44n^3 - 36n^2 - 68n + 54}{12n^2 - 15n},$$

and the solution of the second-order evolution Eq. (111) admits

$$\Delta_r(z) = C_{11}\left(1+z\right)^{\frac{1}{2}(1-\beta)}\text{BesselJ}\left(\frac{\xi}{4}, \frac{2}{3}\frac{\sqrt{3}\pi}{\lambda}\frac{1}{(1+z)^2}\right)$$
$$+C_{12}\left(1+z\right)^{\frac{1}{2}(1-\beta)}\text{BesselY}\left(\frac{\xi}{4}, \frac{2}{3}\frac{\sqrt{3}\pi}{\lambda}\frac{1}{(1+z)^2}\right),$$
$$(112)$$

where $\xi = \sqrt{\beta^2 + 4\gamma - 2\beta + 1}$.

For more clarity, the BesselJ and BesselY presented in Eq. (112) have increasing and decreasing behavior with redshift respectively. For small values of $n$ and $\lambda$, the second terms of the right hand-side Eq. (112) is decreasing with redshift, in other words, increasing with cosmic-time and vice-versa for the first term of this equation.

The solutions of evolution Eq. (112) depend on our free parameters $n$ and $\lambda$. From the definition of (78), we consider $n \geq 0.5$ for numerical plotting and in these intervals of $n$, the value of $\xi$ is always real. Apparently, at $n = 1$ the value of $\bar{\Omega}_r$ becomes unity and $\Omega_T$ reads zero and $\xi = 2\sqrt{2}$, consequently Eq. (112) reduces to radiation dominated case in GR limit, see Eq. (90). For the case of $n \approx 0.5000112$, the value of $\bar{\Omega}_r = \Omega_r \approx 4.48 \times 10^{-5}$ and is closer to the observed value presented in [74]. At $n = 0.5$, $\bar{\Omega}_r = 0$ and $\Omega_T = 1$, here one can say that the torsion fluid is the major component in the system mean that torsion fluid act as a cosmological constant.

In the following plots, we present the numerical results of Eq. (112) for different values of $n$ associated with different wave-lengths, see Figs. 8, 9, 10 and 11 and clearly we see the oscillatory behavior of $\delta(z)$ for the given values of $n$ and $\lambda$.

For $n = 1$, Eq. (111) reduces to GR limit presented in Eq. (89) and the numerical result which is presented in Fig. 10 is exactly the same as $\Lambda$CDM results presented in Fig. 2.

The behavior of the growth of density contrasts is also summarized in Table 2 for short-wavelength mode.

By assuming the same reason as in Sect. 9.1 here, we also consider the generalized power-law $f(T)$ gravity model which is presented in Eq. (102), and the solution of Eq. (111) read as

$$\Delta_r(z) = C_{13}\left(1+z\right)^{\frac{1}{2}(1-\beta_1)}\text{BesselJ}\left(\frac{\xi_1}{4}, \frac{2}{3}\frac{\sqrt{3}\pi}{\lambda}\frac{1}{(1+z)^2}\right)$$
$$+C_{14}\left(1+z\right)^{\frac{1}{2}(1-\beta_1)}\text{BesselY}\left(\frac{\xi_1}{4}, \frac{2}{3}\frac{\sqrt{3}\pi}{\lambda}\frac{1}{(1+z)^2}\right),$$
$$(113)$$

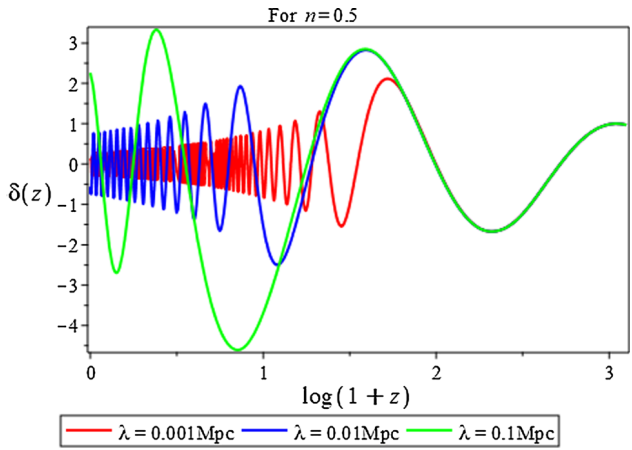

**Fig. 8** $\delta(z)$ versus $z$ for Eq. (112) for short-wavelength mode different $\lambda$ at $n = 0.5$

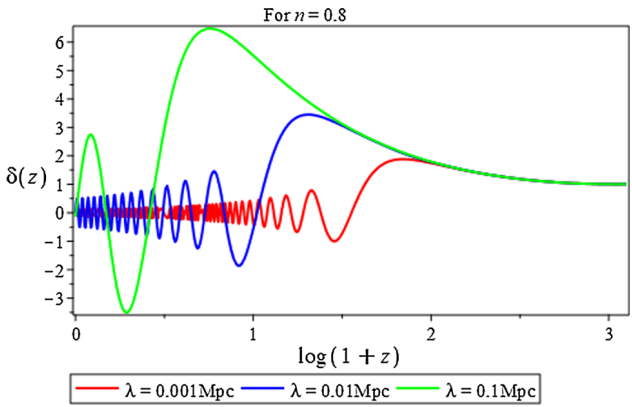

**Fig. 9** $\delta(z)$ versus $z$ for Eq. (112) for short-wavelength mode for different $\lambda$ at $n = 0.8$

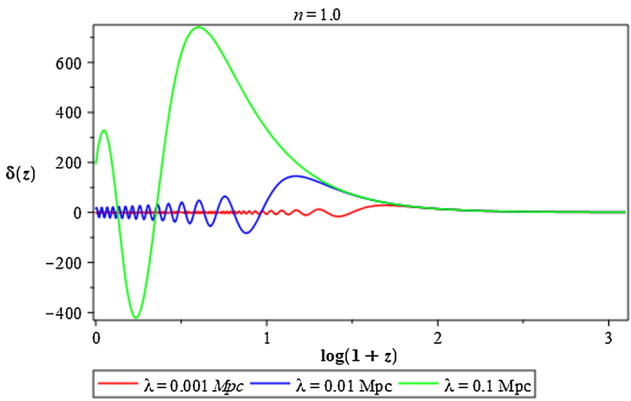

**Fig. 10** $\delta(z)$ versus $z$ for Eq. (112) for short-wavelength mode for different $\lambda$ at $n = 1$

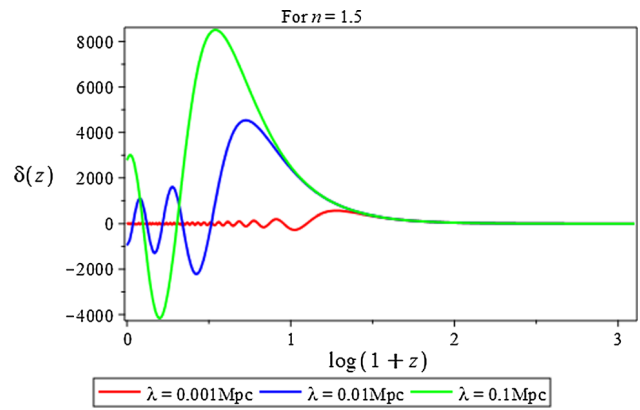

**Fig. 11** $\delta(z)$ versus $z$ for Eq. (112) for short-wavelength mode for different $\lambda$ at $n = 1.5$

**Table 2** We illustrate the features of the density contrast for torsion-radiation system for short-wave length mode and for different ranges of $n$

| Ranges of $n$ | Behavior $\delta(z)$ |
|---|---|
| For $n \geq 0.5$ | Oscillating behavior for all values of $\lambda$ |

The behavior is highly depends only the wave-length range ($\lambda$), and we consider here $n \geq 0.5$ due to the reason of definition (78)

where

$$\beta_1 = \frac{1}{3\,(n\,\Omega_r + n - 1)^2\,(23\,n\,\Omega_r - 24\,\Omega_r - n + 1)}$$
$$\Bigg[ \left( 28\,\Omega_r{}^3 - 2\,\Omega_r{}^2 - 26\,\Omega_r \right) n^4$$
$$+ \left( -69\,\Omega_r{}^3 + 91\,\Omega_r{}^2 + 191\,\Omega_r - 3 \right) n^3$$
$$+ \left( 148\,\Omega_r{}^3 - 87\,\Omega_r{}^2 - 426\,\Omega_r + 9 \right) n^2$$
$$+ \left( -114\,\Omega_r{}^3 - 100\,\Omega_r{}^2 + 383\,\Omega_r - 9 \right) n$$
$$+ 4\,\Omega_r{}^3 + 98\,\Omega_r{}^2 - 122\,\Omega_r + 3 \Bigg],$$

and

$$\xi_1 = \frac{1}{(n\,\Omega_r + n - 1)\,(23\,n\,\Omega_r - 24\,\Omega_r - n + 1)}$$
$$\Bigg[ \left( -16\,\Omega_r{}^2 + 16\,\Omega_r \right) n^3 + \left( 234\,\Omega_r{}^2 - 40\,\Omega_r - 2 \right) n^2$$
$$+ \left( -412\,\Omega_r{}^2 + 28\,\Omega_r + 4 \right) n + 192\,\Omega_r{}^2 - 4\,\Omega_r - 2 \Bigg].$$

By admitting the definition presented in Eq. (81), the numerical results of the density contrast (in this case radiation) for Eq. (113) are presented in the Figs. 12 and 13 for different values of wavelength ranges and $n$. From these figures, we clearly observe that the amplitude of matter density contrast has an oscillatory behavior and the model is viable for $n \ll 1$.

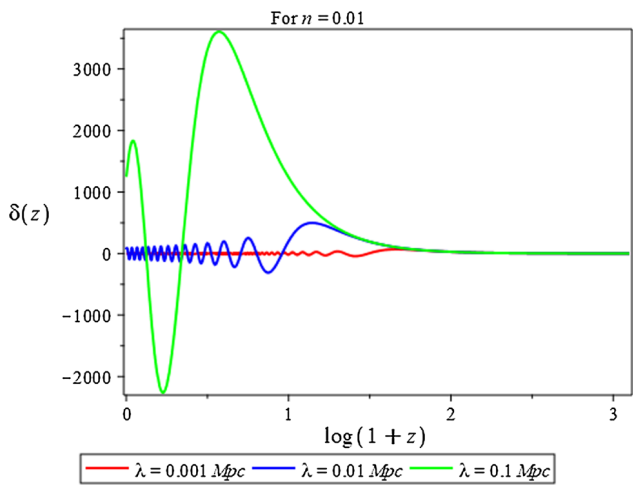

**Fig. 12** $\delta(z)$ versus $z$ for $f(T) = T + \mu T_0(T/T_0)^n$ for short-wavelength mode different $\lambda$ at $n = 0.01$. We use $\Omega_r = 4.48 \times 10^{-5}$ for plotting

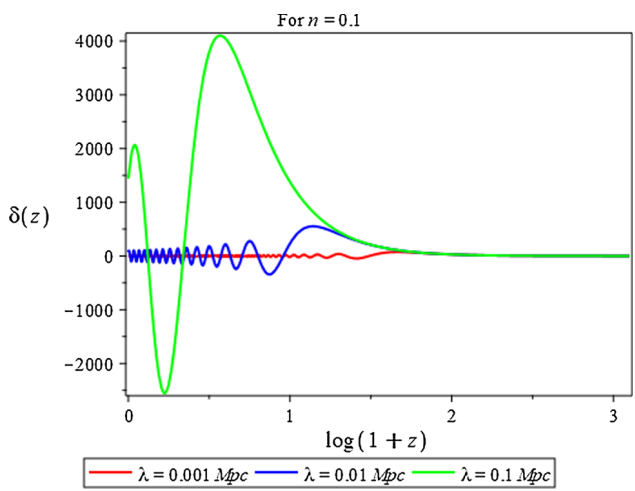

**Fig. 13** $\delta(z)$ versus $z$ for $f(T) = T + \mu T_0(T/T_0)^n$ for short-wavelength mode for different $\lambda$ at $n = 0.1$. We use $\Omega_r = 4.48 \times 10^{-5}$ for plotting

We choose $n = 0.01$ and $0.1$ with $\lambda = 0.001, 0.01$ and $0.1$ in Mpc for illustrative purposes.

In this subsub-section, the density contrasts for $f(T)$ gravity models are studied by using Eq. (111) in torsion-radiation system with short-wavelength mode and its numerical results are presented in Figs. 8, 9, 10 and 11, 12 and 13 for the given $\lambda$ and $n$ values accordingly. The detailed analysis of the growth of matter density fluctuations in torsion-radiation system for short-wavelength mode is made. For instance, the growth of density contrasts in Figs. 8, 9, 10 and 11, 12 and 13 are presented for different values of $n$ at $\lambda = 0.001, 0.01$ and $0.1$ in Mpc. From these figures, we observe that the amplitude of the fluctuations $\delta(z)$ has oscillatory behavior, with amplitudes growing with cosmic-time.

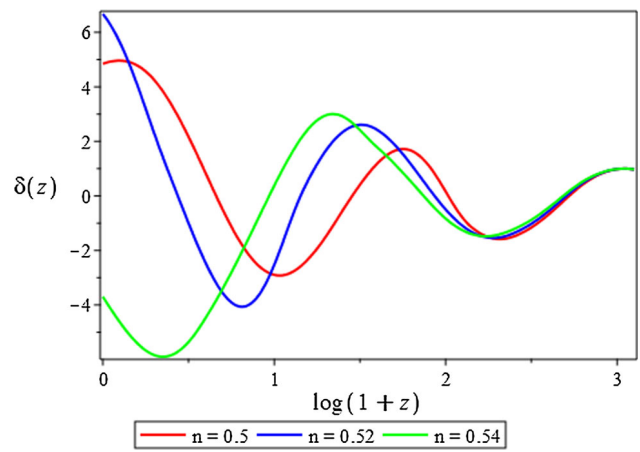

**Fig. 14** $\delta(z)$ versus $z$ in long-wavelength mode for Eq. (115) for the values of $0.5 \leq n \leq 0.54$

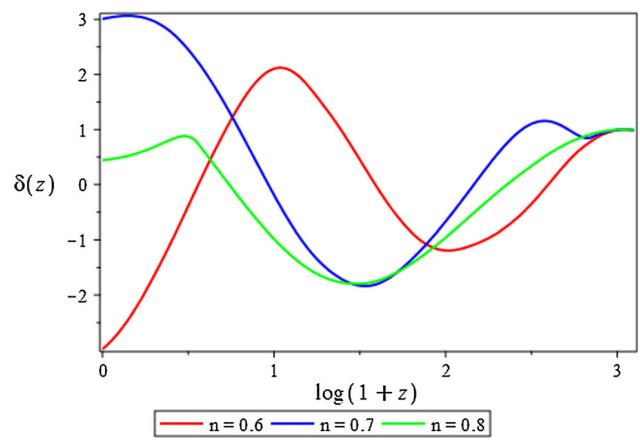

**Fig. 15** $\delta(z)$ versus $z$ in long-wavelength mode for Eq. (115) for the values of $0.6 \leq n < 0.8$

#### 9.2.2 Long-wavelength mode

In the long-wavelength range where $k^2/a^2 H^2 \ll 1$, all cosmological fluctuations begin and remain inside the Hubble horizon. For this limit, our evolution equation (109) takes the form of the Cauchy–Euler equation. So, we apply the same mathematical approach as Eq. (95) on Eq. (111). In this case we have

$$(1+z)^2 \frac{d^2 \Delta_r}{dz^2} + \beta \frac{d\Delta_r}{dz} - \gamma \Delta_r = 0. \tag{114}$$

We assume that $\Delta_r(z) = (1+z)^r$ and we have $r_\pm = \frac{1}{2}(1 - \beta \pm \xi)$. Then, with the $k$-dependence dropped, Eq. (111) admits an exact solution of the form

$$\Delta_r(z) = C_{15} (1+z)^{\frac{1}{2}(1-\beta+\xi)} + C_{16} (1+z)^{\frac{1}{2}(1-\beta-\xi)}. \tag{115}$$

For large values of $n$, the second term of the right-hand side of Eq. (115) is decaying with redshift or growing with cosmic-time and vice-versa for the first term of this equation. The value of $\xi$ highly depends on the $n$ intervals, i.e., $\xi$ is an imaginary for $0.5 < n \leq 0.8$ and real for $n > 0.8$, due to this reason the behavior of $\Delta_r(z)$ has oscillating behavior for $0.5 < n < 0.8$ and growing for $n > 0.8$.

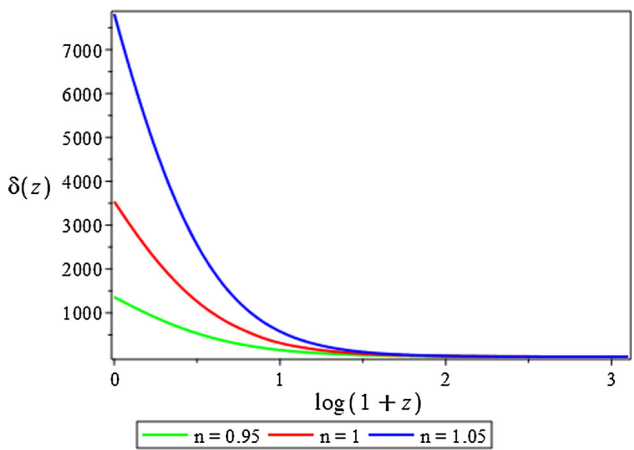

**Fig. 16** $\delta(z)$ versus $z$ in long-wavelength mode for Eq. (115) for the values of $n$ closer to one

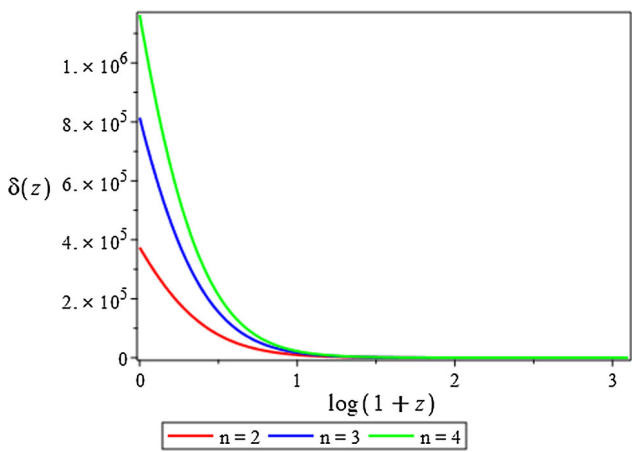

**Fig. 17** $\delta(z)$ versus $z$ in long-wavelength mode for Eq. (115) for the values of $n$ greater than one

In the following, we present the numerical results of Eq. (115) for torsion-radiation system for different values of $n$ in the long-wavelength mode in the Figs. 14, 15, 16 and 17. As mentioned earlier, we clearly observe that the feature of density contrast $\delta(z)$ is very sensitive to the values of $n$. Due to that reason, it has the oscillatory behavior shown in Figs. 14 and 15 for small values of $n$ ($0.5 \leq n \leq 0.8$). For $n$ closer to one and greater than one, the amplitude of $\delta(z)$ is very high and growing with cosmic time or decaying with redshift exponentially (see Figs. 16 and 17), but for $n = 1$ (red solid-line in Fig. 16) the result is recovers GR which is presented in Fig. 3. The main results are summarized in Table 3 for the long-wavelength model.

**Table 3** We illustrate the features of the density contrast for torsion-radiation system for long-wave length mode and for different ranges of $n$

| Range of $n$ | Behavior of $\delta(z)$ |
|---|---|
| $0.5 \leq n \leq 0.8$ | Oscillating |
| $n > 0.8$ | Growing |

The behavior is highly depends only the ranges $(n)$

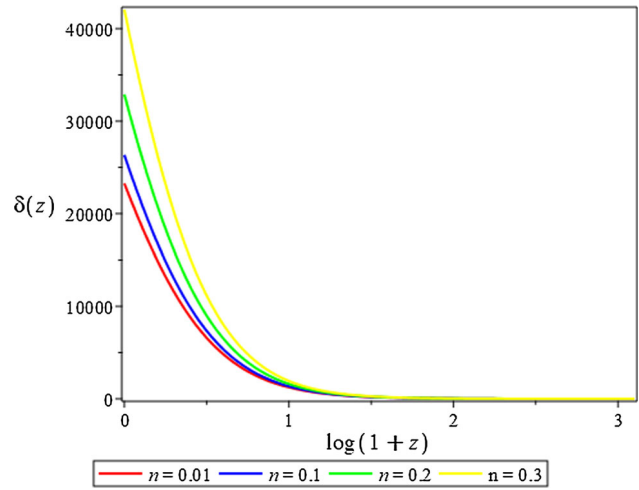

**Fig. 18** $\delta(z)$ versus $z$ in long-wavelength mode for Eq. (116) for the $n \ll 1$

On other-hand, the solutions of Eq. (109) for the generalized power-law $f(T)$ gravity model (102) for the long-wave length limit yields

$$\Delta_r(z) = C_{17}\,(1+z)^{\frac{1}{2}(1-\beta_1+\xi_1)} + C_{18}\,(1+z)^{\frac{1}{2}(1-\beta_1-\xi_1)},$$
(116)

and the numerical results of the density contrasts in this limit are presented in Fig. 18 for $n \ll 1$.

In general, $f(T)$ gravity theory has gained much attention for different cosmological implications and it is shown that the $f(T)$ gravity can be an alternative approach to study the growth of energy density fluctuations for torsion-dust and torsion-radiation systems with $1 + 3$ covariant formalism by applying the paradigmatic power-law $f(T)$ gravity models in Eq. (76) and (102) with power scale factor. We presented the numerical results of Eq. (73), for analyzing the growth of energy density fluctuations from past to present Universe in torsion-dust and torsion-radiation systems for different values of $n$, $\lambda$.

## 10 Conclusions

This work presents a detailed analysis of scalar cosmological perturbations in $f(T)$ gravity theory using the 1+3 covariant gauge-invariant formalism. We defined the gauge-invariant variables and derived the corresponding evolution equations. Then, the harmonic decomposition technique was applied to make the equations manageable for analysis. From that, we obtained exact solutions of the evolution equations for both torsion-radiation and torsion-dust two-fluid systems after considering the quasi-static approximation, and computed the growth of fractional energy density perturbations $\delta(z)$ for the paradigmatic $f(T)$ gravity models and the power-law cosmological scale factor. For the torsion-dust system, we studied the behavior of dust perturbations and observed that $\delta(z)$ is growing with cosmic time. In the torsion-radiation system, we considered short-wavelength and long-wavelength modes. It is observed that the growth of matter density fluctuations for both modes and the density contrast change dramatically for different ranges of $n$ considered, and the amplitude of the density contrasts increases with the values of $\lambda$ and $n$. The density contrasts in our toy $f(T)$ gravity models obviously consistent with GR predictions for $f(T) = T$.

Some of the specific highlights of this work are as follows: in the first model, $f(T) = \mu \left(\frac{T}{T_0}\right)^n$, we have shown the ranges of $n$ for which the perturbation amplitudes $\delta(z)$ oscillate or grow in both dust- and radiation-dominated epochs. For instance, in dust perturbations, the oscillating behavior is observed for $n$ closer to 0.5 while the modes grow for $n > 0.5008$. In radiation perturbations, $\delta(z)$ depicts oscillatory behaviour in the short-wavelength regime for all $n \leq 0.5$, as well as in the long-wavelength regime for $0.5 \leq n \leq 0.8$. But for the range of $n > 0.8$, the amplitudes of $\delta(z)$ grow in the long-wavelength regimes. In the second model, $f(T) = T + \mu T_0(-\frac{T}{T_0})^n$, with the value of $n$ constrained by solar system tests to be in the range $n \ll 1$, we have shown that the amplitude of $\delta(z)$ grows monotonically in the dust-dominated perturbations. For the radiation case, the oscillating behavior is observed in the short-wavelength and the modes monotonically grow for the long-wavelength regime for small values of $n$.

In general, it is evident from our preliminary results that our $f(T)$ models contain a richer set of possibilities whose model parameters can be constrained using up-and-coming observational data and can accommodate currently known features of the large-scale structure power spectrum in the general relativistic and $\Lambda CDM$ limits. We envisage to undertake this aspect of the task for more realistic $f(T)$ models in a multi-fluid cosmological fluid setting in a subsequent work.

**Acknowledgements** SS gratefully acknowledges financial support from Wolkite University and Entoto Observatory and Research Center, Ethiopian Space Science and Technology Institute. JN gratefully acknowledges financial support from the Swedish International Development Cooperation Agency (SIDA) through the International Science Program (ISP) to the University of Rwanda (Rwanda Astrophysics, Space and Climate Science Research Group), project number RWA01. AA acknowledges that this work is based on the research supported in part by the National Research Foundation (NRF) of South Africa with grant numbers 109257 and 112131. AdlCD acknowledges financial support from Project No. FPA2014-53375-C2-1-P from the Spanish Ministry of Economy and Science, Project No. FIS2016- 78859-P from the European Regional Development Fund and Spanish Research Agency (AEI), Project No. CA16104 from COST Action EU Framework Programme Horizon 2020, University of Cape Town Launching Grants Programme and National Research Foundation Grants No. 99077 2016-2018, Ref. No. CSUR150628121624, 110966 Ref. No. BS170509230233, and the NRF Incentive Funding for Rated Researchers (IPRR), Ref. No. IFR170131220846, Blue Skies Programme, Code: 120390Ref.No.BSFP190405427545, National Research Foundation (NRF), South Africa. DFM thanks the Research Council of Norway for their support. The simulations were performed on resources provided by UNINETT Sigma2, the National Infrastructure for High Performance Computing and Data Storage in Norway. This paper is based upon work from the COST action CA15117 (CANTATA), supported by COST (European Cooperation in Science and Technology). SS, JN and AA are grateful for the Institute of Theoretical Astrophysics, University of Oslo, for hosting them during the initial preparation of this manuscript.

**Data Availability Statement** This manuscript has no associated data or the data will not be deposited. [Authors' comment: This manuscript has no associated data or the data will not be deposited.]